\documentclass[showpacs,preprintnumbers,amsmath,amssymb,
superscriptaddress,floatfix]{revtex4}
\usepackage{graphicx}
\usepackage{amsmath}
\usepackage{bm}
\usepackage{mathrsfs}
\usepackage{color}
\usepackage{slashed}
\usepackage{dcolumn}
\usepackage[normalem]{ulem}

\newcommand{\be}{\begin{equation}}
\newcommand{\ee}{\end{equation}}
\newcommand{\ba}{\begin{eqnarray}}
\newcommand{\ea}{\end{eqnarray}}
\newcommand{\nn}{\nonumber}

\newcommand{\rts}{\sqrt{s}}

\begin{document}

\title{The $\phi p$ bound state in the unitary coupled-channel approximation}

\author{Bao-Xi Sun}
\email{sunbx@bjut.edu.cn}
\affiliation{Faculty of Science, Beijing University
of Technology, Beijing 100124, China}

\author{Ying-Ying Fan}
\affiliation{Faculty of Science, Beijing University
of Technology, Beijing 100124, China}

\author{Qin-Qin Cao}
\affiliation{Faculty of Science, Beijing University
of Technology, Beijing 100124, China}

%\date{\today}

\begin{abstract}
The strong attractive interaction of the $\phi$ meson and the proton is reported by ALICE collaboration recently.
The corresponding scattering length $f_0$ is given as $Re(f_0)=0.85\pm0.34(stat)\pm0.14(syst)$fm and $Im(f_0)=0.16\pm0.10(stat)\pm0.09(syst)$fm. The fact that the real part is significant in contrast to the imaginary part indicates a dominate role of the elastic scattering, whereas the inelastic process is less important. In this work, such scattering processes are inspected based on a unitary coupled-channel approach inspired by Bethe-Salpeter equation. The $\phi p$ scattering length is calculated based on this approach, and it is found that the experimental value of the $\phi p$ scattering length can be obtained only if the attractive interaction of the $\phi$ meson and the proton is taken into account. A significant outcome of such attractive interaction is a two-pole structure in the $\phi p$ scattering amplitude. One of the pole, locating at $(1969-i283)$~MeV might correspond to $N(1895)1/2^-$ or $N(1875)3/2^-$ listed in the review of the Particle Data Group(PDG). The other one, locating at ${1949-i3}$~MeV should be a $\phi N$ bound state, which has no counterpart in the PDG data.
\end{abstract}


\maketitle

\section{Introduction}

The strong interaction at low energies, due to its non-perturbative attribute, has led to rich structures of matter as reflected by an abundance of hadronic states. How to understand the nature of those hadrons poses a long-lasting challenge to physicists.
Among the attempts tackling this challenge, chiral perturbation theory provides a systematic approach that effectively accounts for the interaction among hadrons.
By asserting the unitarity condition, the chiral interaction between hadrons results in scattering amplitudes possessing poles in un-physical Riemann sheets in the complex energy plane.
Such poles can be connected with heavier hadronic states as listed in the review of the Particle Data Group\cite{PDG}.
This strategy have made great triumph in interpreting a host of resonances as dynamically generated from the chiral interaction between the baryon and meson ground states, i.e. the spin $1/2$ baryon octet and the pseudoscalar Goldstone-boson octet under the SU(3) flavor symmetry\cite{Kaiser1995, angels, Krippa, Oller:1997ng, ollerulf, carmina, hyodo, Lutz}.
Noticeably, the chiral interaction between the spin $3/2$ decuplet and the Goldstone bosons are also found to be important to understanding some higher resonances \cite{Sarkar:2004jh}.

Besides the above hadronic degrees of freedom,
previous works have shown that vector-meson degrees of freedom may be of crucial significance in understanding the hadron spectra\cite{Molina,Geng,Gonzalez,souravbao,ramos2010,Hosaka2011,Danilkin:2012ua,Guo:2018zvl}.
Remarkably, a recent result of the heavy-ion collision experiment at ALICE indicates an attractive strong interaction potential between the $\phi$ meson and the proton\cite{ALICE}.
But on the other hand, it is troublesome to accommodate those vector-meson degrees of freedom in an effective Lagrangian due to their intermediate masses and their unique decay patterns.
Different models have been proposed for this enterprise \cite{Meissner:1987ge}. Among them, the hidden-gauge symmetry model provides plausible accounts of several important phenomenological laws involving the vector mesons \cite{Meissner:1987ge,hidden1,hidden2,hidden3,Nagahiro}.
Based on this approach, the interaction of vector mesons with baryon octet was studied in detail in Ref.~\cite{ramos2010}.
This work suggested a vanishing elastic potential of the $\phi$ meson with the proton. This conclusion is, however, in contradiction with the recent ALICE result.
From the experiment, the strong interaction between the $\phi$ meson and the proton is shown to be attractive as indicated by the corresponding scattering length as given by the Lednicky-Lyuboshits fit. The real and imaginary parts of the scattering length read respectively,
\ba
\label{eq:length}
{\rm Re}(f_0)&=&0.85\pm0.34\rm (stat)\pm0.14(syst)\,fm, \nn \\
{\rm Im}(f_0)&=&0.16\pm0.10\rm (stat)\pm0.09(syst)\,fm.
\ea
Apparently, the imaginary part of the scattering length is negligible given the error budget. This fact indicates that the elastic scattering is playing a dominant role in the interaction between $\phi$ and the proton, whereas the inelastic scattering turns to be unimportant.
Interestingly, such an attractive potential may be attributed to the $K^+ K^- p$ three-body correlation as suggested by Ref.~\cite{DelGrande:2021mju}.
But for the time being, the dynamical aspects of such a vector-meson-proton system still far away from being comprehended and urgently requires more studies.

The present work is dedicated to a further study on the dynamics of the $\phi$-proton interacting system using an effective Lagrangian approach.
Both the vetor-meson nonet and bayon ground-state octet under flavor SU(3) are incorporated in the effective Lagrangian.
Based on the Lagrangian, we will scrutinise the coupled-channel effects of such a system in more depths.
Especially, we will explore the impacts of an attractive $\phi$-proton potential to the couple-channel dynamics.

In the following section, we will firstly discuss the construction of the effective Lagrangian.  The vector-meson degrees of freedom will be introduced in accordance with the hidden gauge formalism. Next, we will discuss the coupled-channel approach with a restoration of the two-body unitarity. The latter is implemented by a bubble-loop resummation inspired by the Bethe-Salpeter equation. Such a resummation involves a potential which can be obtained directly from the effective Lagrangian.
However, given our current leading-order approach, the effective Lagrangian will not lead to a direct $\phi$-proton potential which should be significant and attractive as suggested by ALICE. We will incorporate such effect by proposing a Yukawa-type potential in addition to the one directly obtained from the Lagrangian. The corresponding discussion is referred to in Section~\ref{sect:Yukawa}.
This work will then end up with an application of our approach to various two-body coupled-channel systems involving the vector-meson nonet and the baryon octet. Special attentions will be paid to a possible resonance dynamically generated in the $\phi N$ scattering channel.

\section{The hidden gauge formalism}
\label{sect:hidden}

The interaction of the vector meson with the octet of baryons can be involved according to the hidden gauge formalism. The  Lagrangian of vector mesons can be written as
\be
\label{eq:vectorLag}
 L=-\frac{1}{4} \langle V_{\mu\nu} V^{\mu\nu} \rangle,
\ee
where the symbol $\langle \rangle$ represents the trace in the $SU(3)$ space and $V_{\mu\nu}$ is the tensor of vector mesons,
\be
V_{\mu\nu} =\partial_\mu V_\nu-\partial_\nu V_\mu-ig \left[ V_\mu, V_\nu \right],
\ee
with $g=\frac{M_V}{2f}$ and the pion decay constant $f=93$MeV\cite{souravbao,ramos2010}.

The Lagrangian of vector mesons in Eq.~(\ref{eq:vectorLag}) supplies an interacting vertex of three vector mesons in the interaction of the vector meson with the octet of baryons, which comes from
\be
\label{eq:VVV}
L^{3V}_{III}=ig\langle \left( V^\mu \partial_\nu V_\mu -\partial_\nu V_\mu V^\mu \right) V^\nu \rangle,
\ee
with the $SU(3)$ matrix of vector mesons
 $V_\mu$ defined by the matrix
\begin{eqnarray}
V_\mu =\frac{1}{\sqrt{2}} \left( \begin{array}{ccc}
\rho^0 + \omega & \sqrt{2}\rho^+ & \sqrt{2}K^{*^+}\\
&& \\
\sqrt{2}\rho^-& -\rho^0 + \omega & \sqrt{2}K^{*^0}\\
&&\\
\sqrt{2}K^{*^-} &\sqrt{2}\bar{K}^{*^0} & \sqrt{2} \phi
\end{array}\right)_\mu.
\end{eqnarray}

Similarly, the Lagrangian for the coupling of vector mesons to the octet of baryons is constructed as
\be
\label{eq:BBV}
L_{BBV}=g\left(\langle \bar{B} \gamma_\mu  \left[V^\mu,B\right]\rangle+ \langle \bar{B} \gamma_\mu B \rangle \langle V^\mu \rangle   \right),
\ee
with the $SU(3)$ matrix of the octet of baryons
\begin{eqnarray}
B =
\left( \begin{array}{ccc}
 \frac{1}{\sqrt{6}} \Lambda + \frac{1}{\sqrt{2}} \Sigma^0& \Sigma^+ & p\\
&& \\
\Sigma^-&\frac{1}{\sqrt{6}} \Lambda- \frac{1}{\sqrt{2}} \Sigma^0 &n\\
&&\\
\Xi^- &\Xi^0 & -\sqrt{\frac{2}{3}} \Lambda
\end{array}\right).
\end{eqnarray}

With the interactions in Eqs.~(\ref{eq:VVV}) and (\ref{eq:BBV}), the $t-$ channel interaction of the vector meson and the baryon can be constructed, where a vector meson is exchanged between them.
Since the scattering process is studied on the energy region near the $\phi$-proton threshold, the momentum of the exchanged vector meson can be neglected, and thus only the mass term is left in the propagator. Therefore, the potential of the vector meson with the octet of baryons takes the form of
\be
\label{eq:202302041614}
V_{ij}=-C_{ij} \frac{1}{4f^2} \left(k^0+k^\prime{}^0 \right) \vec{\epsilon} \vec{\epsilon}^\prime,
\ee
where $k^0(\vec{\epsilon})$ and $k^\prime{}^0(\vec{\epsilon}^\prime)$ are energies(polarization vectors) of the incoming and outgoing vector mesons, respectively, and the $C_{ij}$ coefficient values can be obtained according to the Lagrangian in Eqs.~(\ref{eq:VVV}) and (\ref{eq:BBV}).
Furthermore, due to the mixing of the octet and the singlet of vector mesons in the $SU(3)$ space, the physical states of the $\omega$ and $\phi$ mesons can be written as
\ba
\omega&=&\sqrt{\frac{2}{3}}\omega_1 + \sqrt{\frac{1}{3}}\omega_8, \nn \\
\phi&=&\sqrt{\frac{1}{3}}\omega_1 - \sqrt{\frac{2}{3}}\omega_8.
\ea
Since the singlet state $\omega_1$ does not couple to other vector mesons, and only the octet state $\omega_8$ gives a contribution to the amplitude, the $C_{ij}$ coefficients of the $\omega$ and $\phi$ mesons must be multiplied by factors $\sqrt{\frac{1}{3}}$ and $-\sqrt{\frac{2}{3}}$, respectively.

In the channel of Isospin $I=1/2$, the isospin states for $\rho N$,
$\omega N$, $\phi N$, $K^* \Lambda$, $K^* \Sigma$ can be written as
\begin{equation}
\label{isospin12-rN} |\rho N; \frac{1}{2}, \frac{1}{2} \rangle
~=~-\sqrt{\frac{1}{3}}|\rho_0 p \rangle~-~\sqrt{\frac{2}{3}}|\rho_+
n \rangle,
\end{equation}

\begin{equation}
|\omega N; \frac{1}{2}, \frac{1}{2} \rangle~=~|\omega p\rangle,
\end{equation}

\begin{equation}
|\phi N; \frac{1}{2}, \frac{1}{2} \rangle~=~|\phi p \rangle,
\end{equation}

\begin{equation}
|K^* \Lambda; \frac{1}{2}, \frac{1}{2} \rangle~=~|{K^*}^+ \Lambda
\rangle,
\end{equation}
and
\begin{equation}
\label{isospin12-KS} |K^* \Sigma; \frac{1}{2}, \frac{1}{2}
\rangle~=~\sqrt{\frac{1}{3}}|{K^*}^+ \Sigma^0
\rangle~+~\sqrt{\frac{2}{3}}|{K^*}^0 \Sigma^+ \rangle,
\end{equation}
respectively. We have used the phase convention $\rho^+=-|1,1>$ and
$\Sigma^+=-|1,1>$ for the isospin states in
Eqs.~(\ref{isospin12-rN}) and (\ref{isospin12-KS}), which is
consistent with the structure of the $V^\mu$ and $B$ matrices. It is
apparent that the interaction between isospin states with isospin
orientation $I_z=-1/2$ would generate the same resonances, so only
the isospin states with $I_z=1/2$ are discussed in this section.

The coefficients $C_{ij}$ for the fixed strangeness $S=0$ and isospin $I=1/2$ are listed in Table~\ref{s0i12}.
\begin{table}[htbp]
 \renewcommand{\arraystretch}{1.2}
\centering
\vspace{0.5cm}
\begin{tabular}{l|ccccc}
\hline\hline
 & $\rho N$ & $\omega N$ & $\phi N $  & $K^* \Lambda $ & $K^* \Sigma $\\
 \hline
$\rho N$ & 2 & 0 & 0 & $\frac{3}{2}$ & $-\frac{1}{2}$
  \\
$\omega N$ &  & 0 & 0 & $-\frac{3}{2}\frac{1}{\sqrt{3}}$ &
 $-\frac{3}{2}\frac{1}{\sqrt{3}}$ \\
$\phi N $ &  & & 0 & $-\frac{3}{2}
\left(-\sqrt{\frac{2}{3}}\right)$ &
 $-\frac{3}{2}
\left(-\sqrt{\frac{2}{3}}\right)$ \\
$K^* \Lambda $ &  & & & 0 & 0 \\
$K^* \Sigma $ &  & & & & 2  \\
\hline\hline
\end{tabular}
\caption{
 Coefficients $C_{ij}^{IS}$ for the sector $I=1/2$, $S=0$.
}\label{s0i12}
\end{table}

\section{The unitary coupled-channel approach}
\label{sect:BS}

A full scattering amplitude can be expressed in an integral representation according to Bethe-Salpeter equation. By applying the on-shell condition to the potential $V$ involved, the equation can be reduced to a resummation of bubble-loop series and we end up with the scattering amplitude of $\phi p\rightarrow \phi p$, reading as
\be
T=\left[1-VG \right]^{-1} V, \label{eq:BS}
\ee
with $G$ being the bubble-loop function of the intermediate stable vector meson and baryon.
According to this approach, coupled-channel effects are well incorporated.
In the dimensional regularization scheme, the loop function $G$ takes the form of
\begin{eqnarray}
\label{eq:g-function}
  G_i(\rts) &=&  \frac{2 M_i}{(4 \pi)^2}
  \left\{
        a_i(\mu) + \log \frac{m_i^2}{\mu^2} +
        \frac{M_i^2 - m_i^2 + s}{2s} \log \frac{M_i^2}{m_i^2}
  \right.
  \\
     &+& \frac{Q_i(\rts)}{\rts}
    \left[
         \log \left(  s-(M_i^2-m_i^2) + 2 \rts Q_i(\rts) \right)
      +  \log \left(  s+(M_i^2-m_i^2) + 2 \rts Q_i(\rts) \right)
    \right.
  \nonumber
  \\
  & &
  \Biggl.
    \left.
      - \log \left( -s+(M_i^2-m_i^2) + 2 \rts Q_i(\rts) \right)
      - \log \left( -s-(M_i^2-m_i^2) + 2 \rts Q_i(\rts) \right)
    \right]
  \Biggr\},
  \nonumber
\end{eqnarray}
where the loop function follows the form in Ref.~\cite{ollerulf}, and $\mu=630$MeV and $a_i(\mu)=-2.0$ in the calculation,
and the two-body unitarity is therefore preserved in \eqref{eq:BS}.

Since the vector mesons, particularly the $\rho$ and the $K^*$, are rather
broad, a proper account for their widths is required when performing the loop integral. We follow Ref.~\cite{ramos2010}
and convolute the vector meson-baryon loop function with the mass distribution of such a meson, i.e., by replacing the $G$ function appearing in
Eq.~(\ref{eq:g-function}) by $\tilde{G}$
\begin{eqnarray}
\tilde{G}(s)&=&\frac{1}{N}\int\limits^{(m_1+2\Gamma_1)^2}\limits_{(m_1-2\Gamma_1)^2}
d\tilde{m}^2_1\left(-\frac{1}{\pi}\right)\mathrm{Im}\frac{1}{\tilde{m}^2_1-m_1^2+i m_1 \tilde{\Gamma}_1 (\tilde{m_1})}\nonumber\\
\nonumber\\
&&\times G(s,\tilde{m}_1^2,M^2)
\end{eqnarray}
with
\begin{eqnarray}
N&=&\int\limits^{(m_1+2\Gamma_1)^2}\limits_{(m_1-2\Gamma_1)^2}
d\tilde{m}^2_1\left(-\frac{1}{\pi}\right)\mathrm{Im}\frac{1}{\tilde{m}^2_1-m_1^2+i m_1 \tilde{\Gamma}_1 (\tilde{m_1})}\nonumber\\
\end{eqnarray}
where $m_1$  and $\Gamma_1$ are the mass and
width of the vector meson in the loop. We only take into
account the widths of the $\rho$ and the $K^*$. In the case of the
$\omega$ or $\phi$, one or both of the kernels of these integrals
will reduce to a delta function $\delta(\tilde{m}^2-M^2)$. The
$\tilde{\Gamma}_i$ function is energy dependent and has the form of
\begin{equation}
\tilde{\Gamma}(\tilde{m})=\Gamma_0  \frac{m^2}{\tilde{m}^2}  \frac{q^3_\mathrm{off}}{q^3_\mathrm{on}}\Theta(\tilde{m}-m_1-m_2)
\end{equation}
with
\begin{equation}
q_\mathrm{off}=\frac{\lambda^{1/2}(\tilde{m}^2,m_\pi^2,m_\pi^2)}{2\tilde{m}},\quad
q_\mathrm{on}=\frac{\lambda^{1/2}(M_\rho^2,m_\pi^2,m_\pi^2)}{2 M_\rho},
\end{equation}
$m_1=m_2=m_\pi$, and $m=m_\rho$ for the $\rho$ or
\begin{equation}
q_\mathrm{off}=\frac{\lambda^{1/2}(\tilde{m}^2,m_K^2,m_\pi^2)}{2\tilde{m}},\quad
q_\mathrm{on}=\frac{\lambda^{1/2}(M_{K^*}^2,m_K^2,m_\pi^2)}{2 M_{K^*}},
\end{equation}
$m_1=m_\pi$, $m_2=m_K$ and $m=m_{K^*}$ for the $K^*$, where $\lambda$ is the
K\"allen function, $\lambda(x,y,z)=(x-y-z)^2-4yz$, and $\Gamma_0$ is the
nominal width of the $\rho$ or the $K^*$.

The scattering amplitude $T_{ij}$ takes a form of
\be
T_{ij}=\frac{g_i g_j}  {\sqrt{s}-\sqrt{s_0}}
\ee
at the pole position $\sqrt{s_0}$ in the complex energy plane approximately. Therefore the coupling to the $i$th channel $g_i$ can be calculated with the residue of the scattering amplitude $T_{ii}$ at the pole position. After $g_i$ is obtained, the couplings to the other channels $g_j$ are easy to be evaluated with the residues of inelastic scattering amplitudes $T_{ij}$.

When the Bethe-Salpeter equation is solved in the unitary coupled-channel approximation, the vertices of the baryon octet and vector mesons are used, and the outlines in Feynman diagrams are cut off, which implies the polarization vector of mesons is eliminated in the calculation. More detailed discussions on this question can be found in Ref~\cite{Sun2017}. In the present work, we adopt the original formula in Refs.~\cite{Gonzalez,souravbao,ramos2010}, where $\vec{\epsilon} \vec{\epsilon}^\prime=-1$ in Eq. (\ref{eq:202302041614}) is assumed. Actually, this scheme is consistent with the formula discussed in in Ref~\cite{Sun2017}. As for the spin of the nucleon, the interaction would vanish if the spin orientations of the initial and final nucleons are different from each other. Therefore, the vertices used in the Bethe-Salpeter equation are independent on the spin of the system. At this point, it is different from the case in the chiral SU(3) quark model\cite{Huang:phiN}.

\section{The Yukawa-type potential of the $\phi$ meson with the proton }
\label{sect:Yukawa}

The Yukawa-type potential of the $\phi$ meson with the proton takes the form of
\be
\label{eq:Yukawa}
V(r)=-A \frac{\exp(-\alpha r)}{r},
\ee
with $A=0.021\pm0.009\pm0.006(syst)$ and $\alpha=65.9\pm38.0(stat)\pm17.5(syst)$MeV\cite{ALICE}.
Therefore, the $\phi$- proton potential in the momentum space can be obtained with a Fourier transformation of Eq.~(\ref{eq:Yukawa}), which is written as
\be
\label{eq:Yukawamomen}
V(\vec{q})=\frac{-g^2_{\phi N}}{\vec{q}^2+\alpha^2},
\ee
with $\frac{g^2_{\phi N}}{4 \pi}=A$.
Apparently, the potential in Eq.~(\ref{eq:Yukawamomen}) is a non-relativistic form supposing that the three-momentum of the proton(phi-meson) is far lower than the proton(phi-meson) mass. In this case, the zero component of the momentum transfer in the phi-proton $t-$channel interaction tends to zero in the non-relativistic approximation. More detailed iterations can be found in the appendix part of Ref.~\cite{sun:Fermi}. It should be noticed that the phi-proton coupling in this work $g_{\phi N}^2/4\pi=A$ is different from that in Ref.~\cite{ALICE}, where $g_{\phi N}^2=A$.

When the Bethe-Salpeter equation is solved, the Yukawa type potential in Eq.~(\ref{eq:Yukawamomen}) is multiplied by a mass of $\phi$ meson.
and the three-momentum transfer $\vec{q}$ vanishes at the $\phi$- proton threshold. Therefore,
\be
\label{eq:Yukawamomenzero}
V(\vec{0})=\frac{-g^2_{\phi N} m_\phi}{\alpha^2},
\ee
which can be used as the potential involved in the Bethe-Salpeter resummation.

\begin{figure}
\includegraphics[width=0.6\textwidth]{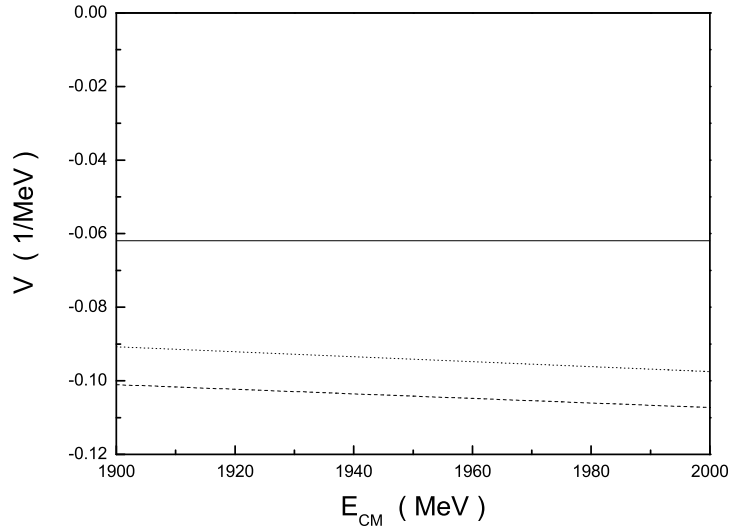}
\caption{
The potential of the vector meson and baryon octet as a function of the total
energy of the system $E_{CM}$ for elastic processes with
strangeness $S=0$ and isospin $I=1/2$.
The $\phi N$, $\rho N$ and $K^* \Sigma$ cases
are represented by solid, dash and dot lines, respectively, while the $\omega N$ and $K^* \Lambda$ potentials are zero.
}
\label{fig:potential}
\end{figure}

In the calculation of the present work, we set $A=0.021$ and $\alpha=65.9$MeV, and the corresponding potentials of the vector meson and baryon octet are depicted in Fig.~\ref{fig:potential}. There are not direct interactions of $\omega N$, $\phi N$ and $K^* \Lambda$ when the hidden gauge symmetry is taken into account. However, the attractive $\phi N$ potential in Eq.~(\ref{eq:Yukawamomenzero}) is a constant and about half of the $\rho N$ potential, and it is no doubt that it would play a critical role in  the sector of strangeness $S=0$ and isospin $I=1/2$.

In Ref.~\cite{ALICE}, it is assumed that the elastic process of the $\phi$ meson and the proton plays a dominant rule, while the inelastic process is not important.

\section{The $\phi N$ bound state}
\label{sect:bound}

With the coefficients listed in Table~\ref{s0i12}, the scattering amplitude $T$ of the $\phi N\rightarrow \phi N$ reaction is evaluated by solving the Bethe-Salpeter equation in the unitary coupled-channel approximation, and then the $\phi N$ scattering length can be obtained according to the $\phi N$ elastic scattering amplitude at the thresold, i.e.,
\be
a_{\phi N}=\frac{M_N}{8\pi \sqrt{s}}T_{\phi N\rightarrow \phi N}(\sqrt{s}=m_\phi+M_N).
\ee
With $\mu=630$MeV and $a_l(\mu)=-2.0$, the real part of the $\phi N$ scattering length is about $-0.15$fm, and the imaginary part is $0.11$fm approximately.
Apparently, the $\phi N$ scattering length takes a small value and can not be consistent with the value announced by ALICE collaboration recently, as given in Eq.~(\ref{eq:length}). Even if the subtraction constants are adjusted, it is still far away from the experimental value.

In order to obtain the $\phi N$ scattering length consistent with the experimental value supplied by the ALICE collaboration, we include the attractive $\phi N$ potential as given by Eq.~(\ref{eq:Yukawamomenzero}) when performing the Bethe-Salpeter resummation.
By adjusting the subtraction constant values, the $\phi N$ scattering length is obtained, and it is consistent with the experimental value in the range of uncertainty. The subtraction constant values and the corresponding $\phi N$ scattering length are listed in Table~\ref{table:subS0I12}.

\begin{table}[htbp]
\begin{tabular}{ccccc|cc}
\hline\hline
       $a_{\rho N}$&$a_{\omega N}$&$a_{\phi N}$&$a_{K^* \Lambda}$&$a_{K^* \Lambda}$ & $f_0$ \\
\hline
        -2.0  &  -2.0  &  -2.4  &  -1.9  &  -1.8  &  0.86+i0.19  \\
\hline \hline
\end{tabular}
\caption{The subtraction constants and the real and imaginary parts of the corresponding $\phi N$ scattering length obtained in unitary coupled-channel approximation when the $\phi N$ attractive interaction is taken into account, where the scattering length $f_0$ is in units of MeV, and the regularization scale $\mu=630$MeV.} \label{table:subS0I12}
\end{table}

\begin{figure}
\includegraphics[width=0.6\textwidth]{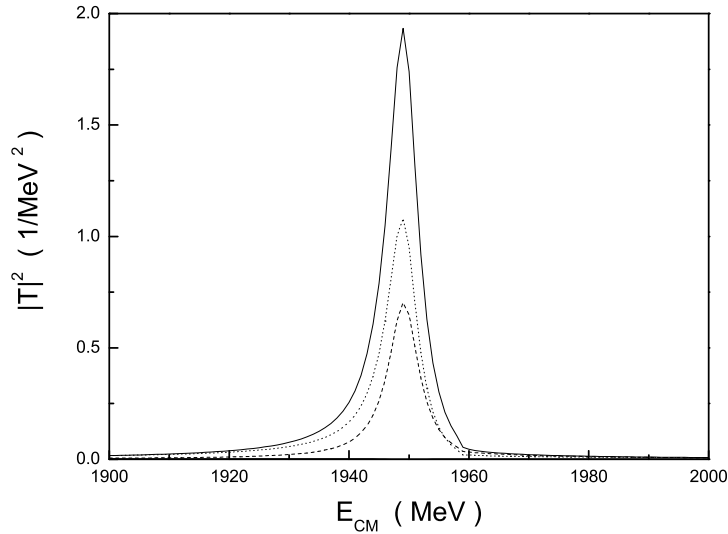}
\caption{
The amplitude squared $|T|^2$ as a function of the total
energy of the system $E_{CM}$ for elastic processes with
strangeness $S=0$ and isospin $I=1/2$.
The $\phi N$, $K^* \Lambda$ and $K^* \Sigma$ cases
are represented by solid, dash and dot lines, respectively, while the amplitudes squared of $\rho N$ and $\omega N$ are small and the corresponding curves are invisible in this figure.
}
\label{fig:s0i12}
\end{figure}

With the subtraction constants listed in Table~\ref{table:subS0I12}, the scattering amplitudes $|T_{ii}|^2$ of the vector meson and the baryon octet are calculated in the unitary coupled-channel approximation. In Fig.~\ref{fig:s0i12}, the scattering amplitude with strangeness $S=0$ and isospin $I=1/2$ is displayed. In this case a peak around 1950MeV is visible in the $\phi N$, $K^* \Lambda$ and $K^* \Sigma$ channels. Since it is below the $\phi N$ threshold, which could be associated with a $\phi N$ bound state.
Actually, there is another pole of $|T_{ii}|^2$ being detected around 1970MeV in the $K^* \Lambda$ and $K^* \Xi$ channel. However it is far from the real axis, and therefore leads to no visible effects in Fig.~\ref{fig:s0i12}. Since it is higher than the $\phi N$ threshold, it can be regarded as a $\phi N$ resonance.

The pole positions and corresponding couplings of these two states to the different channels, obtained from the residues
at the poles are shown in Table~\ref{table:coupling}.
Since the inelastic scattering amplitudes are far lower than the elastic ones, all couplings are calculated according to elastic processes.
It can be seen that the state at $1949-i3$MeV couples strongly to the $\phi N$ channel, and the state at $1969-i283$MeV mainly couples to the $K^* \Lambda$ and $K^* \Sigma$ channels.

\begin{table}[htbp]
\begin{tabular}{c|c|c|c|c|c}
\hline\hline
 Pole positions   &  $\rho N$          &  $\omega N$   &   $\phi N$          & $K^* \Lambda$ & $K^* \Sigma$ \\
\hline
$1949-i3$   & $0.0+i0.0$  &  $0.1+i0.0$ &  $2.1+i0.1$ & $1.6+i0.3$ & $1.8+i0.0$  \\
$1969-i283$ & $0.1+i0.1$  &  $0.0+i0.2$ &  $0.1-i0.1$  & $0.3-i0.4$   & $0.2-i0.0$  \\
\hline \hline
\end{tabular}
\caption{Pole positions in units of MeV and coupling constants to various channels
of resonances with strangeness $S=0$ and isospin $I=1/2$.}
\label{table:coupling}
\end{table}

The Yukawa type potential of the phi-proton system is only valid when the total energy of the system is close to the phi-proton threshold. At this point, the three-momentum transfer in the phi-proton potential is not large and can be neglected in the calculation, and a phi-nucleon bound state is found when the scattering amplitude is analyzed. Since the bound state is about 10MeV lower than the phi-nucleon threshold, it is reasonable to neglect the three-momentum transfer in the phi-proton potential in Eq.~(\ref{eq:Yukawamomen}) when the Bethe-Salpeter equation is solved.

Actually, we also studied the case that the three-momentum transfer $\vec{q}$ is not zero.
By solving the Bethe-Salpeter equation, it is found that two poles appear at $1949-i3$~MeV and $1970-i283$~MeV on the complex energy plane, respectively. It is apparent that these results are almost the same as those with $\vec{q}=0$, where the three-momentum transfer is neglected. Moreover, the same value of the phi-nucleon scattering length is obtained with the subtraction constants listed in Table \ref{table:subS0I12}. The coupling constants to different channels are calculated again, and it shows that the values of them change slightly. However, these two states still couple strongly to the $\phi N$, $K^* \Lambda$, $K^* \Sigma$ channels.

In the S-wave approximation, two poles of the scattering amplitude with strangeness $S=0$ and isospin $I=1/2$ are found in the unitary coupled-channel approximation, and they are degenerate in $J^P=1/2^-,3/2^-$.
The resonance state at $1969-i283$MeV might correspond to either the baryon $N(1895)1/2^-$ or $N(1875)3/2^-$ particles, while the pole at $1949-i3$MeV has no counterpart in the PDG data since its decay width is too small. It might be a $\phi N$ bound state caused by the strong attractive interaction between the $\phi$ meson and the proton.
The pole position and corresponding PDG data are listed in Table~\ref{table:PDGdata}.

\begin{table}[htbp]
\begin{tabular}{c|ccccc}
\hline\hline
 Pole position          &name &  $J^P$ & status & mass         & width    \\
\hline
 $1949-i3$                &  -     &  -      &   -       &    -            & - \\
\hline
 $1969-i283$              & $N(1895)$      &   $1/2^-$ &****    &   1870-1920       &     80-200           \\
                          & $N(1875)$      &   $3/2^-$ & ***    &   1850-1920       &     120-250          \\
 \hline \hline
\end{tabular}
\caption{The pole position and corresponding PDG data with strangeness $S=0$ and isospin $I=1/2$, all in units of MeV
.} \label{table:PDGdata}
\end{table}

The pole position in the complex energy plane is closely related to subtraction constants when the Bethe-Salpeter equation is solved.
With the subtraction constant $a=-2$ for variant channels, a pole at $1979-i56$MeV is generated in the sector of strangeness $S=0$ and isospin $I=1/2$, which is similar to the result in Ref.~\cite{ramos2010}. However, in this work, the subtraction constants are determined in order to obtain the experimental phi-nucleon scattering length, as listed in Table~\ref{table:subS0I12}. Therefore, a pole at $1969-i283$MeV is generated, which is higher than the threshold of the phi-meson and the nucleon, and can be regarded as a phi-N resonance state. While the particles $N(1895)1/2^-$ and $N(1875)3/2^-$ is under the phi-N threshold, might be bound states of the phi-meson and the nucleon.

The discrepancy of the calculation results from the experimental data might come from the SU(3) hidden gauge symmetry, by which the vector mesons are included in the interaction Lagrangian. Anyway, the masses of vector mesons are different from each other, so the SU(3) hidden gauge symmetry are only reliable approximately. Moreover, perhaps some high-order corrections on the kernel should be taken into account when the Bethe-Salpeter is solved in the unitary coupled-channel approximation. These topics would be investigated continuously in the future work of ours.

The $\phi$-nucleon bound state is also studied in the chiral SU(3) quark model\cite{Huang:phiN}, where the exchange of scalar meson $\sigma$ is dominant in the attractive potential of the phi-meson and the nucleon.
In the phi-nucleon potential in Eq.~(\ref{eq:Yukawamomen}), the three-momentum transfer is negligible in the calculation, and thus only the ratio
$g^2_{\phi N}/\alpha^2 $ is critical in the calculation, which equals to $6.0\times 10^{-5}$MeV$^{-2}$ approximately. In the chiral SU(3) quark model,
the attractive potential via a scalar meson can be obtained when $g_{\phi N}$ and $\alpha$ in Eq.~(\ref{eq:Yukawamomen}) are replaced with the coupling constant $g_{ch}$ and the mass of the scalar meson $\sigma$ in
Ref.~\cite{Huang:phiN}, respectively.
In the region of the phi-nucleon threshold, the three-momentum transfer can be neglected and the value of $g^2_{ch}/m_\sigma^2 \approx 2.4\times 10^{-5}$MeV$^{-2}$ is obtained in the extended chiral SU(3) quark model of type II, which has the same order of magnitude as that of $g^2_{\phi N}/\alpha^2 $ in this work. Moreover, the bound state is about 10MeV lower than the phi-nucleon threshold, and it is consistent with the binding energy of the phi-nucleon bound state obtained in the chiral SU(3) quark model\cite{Huang:phiN}.

\section{Summary}
\label{sect:summary}

In the present work, we investigate the implications of a strong attractive potential of the  $\phi$ meson and proton as recently reported by the ALICE collaboration.
An effective description of such a potential is obtained by introducing a Yukawa-type formalism on top of the vector-meson-nonet and baryon-octet interaction given by the effective Lagrangian.
We deliberate on the impacts of such a potential to the coupled-channel effects of vector-meson and baryon scattering systems.
A unitary approach has been employed to implement the coupled-channel effects. This approach involves a series resummation reduced from Bethe-Salpeter equation, and we arrive at a $\phi N$ scattering length consistent with the experiment.
Based on such potential, we investigate the coupled-channel effect, and
special attentions have been paid to the coupled-channel effects of the vector-meson and baryon scattering system with strangeness $S=0$ and isospin $I=1/2$.
Two poles have been found in the complex energy plane of the corresponding scattering amplitude.
One pole locates at  $\sqrt{s}=(1969-i283)$MeV, and might correspond to either the baryon $N(1895)1/2^-$ or $N(1875)3/2^-$.
In addition, we predict a possible $\phi N$ bound state, which has not been found in the PDG. This state leads to another pole structure at $\sqrt{s}=(1949-i3)$MeV in the scattering amplitude.

\begin{acknowledgments}
Bao-Xi Sun would like to thank Xiao-Yu Guo for reading and revising the whole manuscript. Moreover, we appreciate the constructive and helpful advice from referees of Communications in Theoretical Physics.
\end{acknowledgments}

\newpage

\end{document}